\documentstyle[aps,twocolumn,epsfig]{revtex}
\begin{document}
\newcommand{\ve}[1]{\mbox{\boldmath $#1$}}
\twocolumn[\hsize\textwidth\columnwidth\hsize
\csname@twocolumnfalse%
\endcsname

\draft

\title{Lieb Mode in a Quasi One-Dimensional Bose-Einstein
Condensate of Atoms}
\author{A. D. Jackson$^1$ and G. M. Kavoulakis$^2$}
\date{\today}
\address{$^1$Niels Bohr Institute, Blegdamsvej 17, DK-2100 Copenhagen \O,
        Denmark, \\
         $^2$Mathematical Physics, Lund Institute of Technology, P. O. Box 118,
           S-22100 Lund, Sweden}
\maketitle

\begin{abstract}

We calculate the dispersion relation associated with a solitary wave
in a quasi-one-dimensional Bose-Einstein condensate of atoms confined 
in a harmonic, cylindrical trap in the limit of weak and strong
interactions. In both cases, the dispersion relation is linear for 
long wavelength excitations and terminates at the point where the
group velocity vanishes. We also calculate the dispersion relation
of sound waves in both limits of weak and strong coupling.

\end{abstract}
 
\vskip0.0pc]

Bose-Einstein condensates of trapped alkali-metal atoms \cite{PS}
offer a rich source of interesting non-linear phenomena.  At mean-field 
level, the effects of atom-atom interactions can be described as a one-body 
potential proportional to the local density of atoms.  The order parameter, 
i.e., the condensate wavefunction, then satisfies a non-linear Schr\"odinger 
equation which also includes the effect of the confining (harmonic) potential.

Many authors have discussed the properties of solitary waves
in a Bose-Einstein condensate of trapped alkali-metal atoms
\cite{MBB,SBB,ADP,JKP,MHS,BA,Feder,WLN}. Solitary waves were created and
studied experimentally by Burger {\it et al}.\ \cite{H} and by Denschlag 
{\it et al}.\ \cite{N}.  Solitary waves result as a balance between the 
energy cost associated with the Heisenberg principle, $\hbar^2/2 m \xi^2$,
where $m$ is the atom mass and $\xi$ is the characteristic length 
of the solitary wave, and the energy gain due to the local density
variation of the system, which is of order $n U_0$. Here $n$ is
the atomic density, and $U_0=4 \pi \hbar^2 a/m$ is the effective two-body
interaction matrix element, with $a$ being the scattering
length for atom-atom collisions. In the present study, we consider
repulsive interactions with $a > 0$.  Equating these two terms,
one sees that the characteristic size of a solitary wave is set by 
the coherence length $\xi$ which satisfies the equation $\hbar^2/ 2 m \xi^2 = 
n U_0$. In the experiments of Refs.\cite{H,N}, the effective interaction 
between the atoms was repulsive, and the solitary waves were thus density 
depressions.  For an attractive effective interaction, the solitary waves 
are expected to be elevations in the density.

Since the atoms are confined, momentum is not a good quantum
number. However, it is possible to use cigar-shaped traps which are 
very long along the $z$ axis. These systems are quasi one-dimensional 
\cite{JKP,KP}, and the momentum along this axis is approximately a good 
quantum number.  

An interesting question arises in this context.  Some 40 years ago, 
Lieb considered a purely one-dimensional Bose gas of atoms interacting
via a contact potential and predicted two distinct modes of excitation
\cite{Lieb}.  One was identified as the usual Bogoliubov mode.  The other 
class of excitations was later shown by Kulish {\it et al.} \cite{KMF} 
to be associated with solitary waves (see also Ref.\,\cite{J}.) 
These authors demonstrated that the dispersion relation resulting from 
solitary wave excitation is associated to that predicted by Lieb.
It is thus reasonable to ask (at least in the case of quasi one-dimensional
atomic condensates) whether this ``Lieb mode'' exists. Actually, in the
recent study of Ref.\,\cite{KNSQ} the Lieb mode was examined in one
dimension. Although the theoretical prediction for this mode seems firm, it 
has never been observed experimentally.  In this regard, it is interesting 
that Stamper-Kurn {\it et al.}\,\cite{Bragg} and Ozeri 
{\it et al.}\,\cite{OSKD} have recently managed 
to probe the long wavelength phonon spectrum associated with the Bogoliubov 
mode in a cigar-shaped condensate of atoms using Bragg spectroscopy.
As we argue below, the Lieb mode should be present in such a system and 
may be observable. Since for long-wavelength excitations the Lieb mode
coincides with the usual Bogoliubov mode of sound waves, it is crucial that 
the momentum imparted to the cloud be appropriately large for the two modes to
have distinct energies. One, for example, could excite the cloud using the
method of phase imprinting, in order to create a solitary wave, and then
measure the excitation energy and the corresponding momentum.

In the present study we derive the dispersion relation associated
with the Lieb mode. Reference \cite{KPa} (and, recently, Ref.\,\cite{KPab})
has dealt with the same problem for a different range of parameters using a
full three-dimensional numerical calculation based on the nonlinear
Gross-Pitaevskii equation. In this calculation, the solitary wave is found to
be a hybrid between a one-dimensional soliton and a three-dimensional vortex
ring, but the method is applicable in our limit, as well. 
Since the present results deal with weaker interactions, we have 
chosen to adopt a description which neglects the contribution of vortex rings.
We distinguish between two limits.  In the limit of weak interactions, 
$n_0 U_0 \ll \hbar \omega_{\perp}$, where $n_0$ is the maximum density of
atoms far away from the wave, and $\omega_{\perp}$ is the frequency of the
trapping potential transversely to the long axis of the trap,
the resulting equation is the ordinary nonlinear Gross-Pitaevskii equation.
In the opposite limit of strong interactions $n_0 U_0 \gg \hbar
\omega_{\perp}$, the resulting equation is a modified Gross-Pitaevskii
equation, in which the nonlinear term is proportional to the magnitude of the 
order parameter. Finally, we calculate the usual Bogoliubov mode in both 
regimes of weak and strong interactions and comment on the limits of validity 
of our study. 

{\it Model.}
We start with a $T=0$ Bose-Einstein condensate of atoms confined in a 
cylindrical harmonic potential, $V = m \omega_{\perp}^2 (x^2+y^2)/2$ and 
assume wave motion along the $z$ axis.  There is no confinement along the 
$z$ axis, and away from the wave, there is a uniform density of atoms, 
$n(x,y)$, which is independent of $z$.  For a short-ranged atom-atom 
interaction, $V_{\rm int}({\bf r}-{\bf r}')=U_0 \delta({\bf r}-{\bf r}')$,
the Gross-Pitaevskii equation for the order parameter $\Psi$
has the form
\begin{eqnarray}
   i \hbar \, \partial_t \Psi =
  (-\hbar^2 {\bf \nabla}^2 / {2m} + U_0 |\Psi|^2 + V) \Psi.
\label{GPequ}
\end{eqnarray}
Following Ref.\cite{KP}, we assume that the transverse dimension of the
cloud is sufficiently small and the corresponding time scale sufficiently 
rapid that the transverse profile of the particle density can adjust to 
the equilibrium form appropriate for the instantaneous number of particles 
per unit length.  The problem becomes one-dimensional, and the solitary pulse 
can be described by a local velocity, $v(z)$, and a local density of
particles per unit length, $\sigma(z)$ \cite{KP},
$\sigma(z)=\int dx dy \, |\Psi(x,y,z)|^2$.
With this assumption, the wavefunction may be written in the form 
$\Psi({\bf r},t)=f(z,t) \, g(x,y,\sigma)$ \cite{JKP},
where $g$ is the equilibrium wavefunction for the transverse motion which 
depends on time implicitly through the time dependence of $\sigma$.  We 
choose $g$ to be normalized so that $\int |g|^2 dx dy = 1$ and thus, from the 
equations above, $|f|^2=\sigma$. 

To proceed, we consider two opposite limits, namely the weak-coupling 
limit and the Thomas-Fermi regime. The transition between the two
limits occurs for $\sigma_0 a \sim 1/4$ \cite{JKP}, where $\sigma_0$ is the 
value of $\sigma$ far away from the wave. 

{\it Weak-coupling limit.}
We first consider the weak coupling regime.  Although 
this has traditionally been an academic limit, it is now possible to create 
Bose-Einstein condensates in cigar-shaped traps \cite{AK} which realize 
the weak-interaction limit.  In this case $|g|^2$ has a Gaussian form,
$|g|^2 = (\pi a_{\perp}^2)^{-1} e^{-({\rho}/a_{\perp})^2}$.
As shown in Ref.\cite{JKP}, $f$ satisfies the equation 
\begin{eqnarray}
 i \hbar \, \partial_t f =
  - (\hbar^2/2m) \partial_z^2 f +
 \hbar \omega_{\perp} (1+ 2 a |f|^2) f.
\label{fgp1w}
\end{eqnarray}
We see from this equation that $f \propto e^{-i \omega_{\perp} 
(1+2a \sigma_0) t}$ as $|z| \to \infty$.  Thus, we rewrite Eq.\,(\ref{fgp1w}) 
using the variable $w=f e^{i \omega_{\perp} (1+2a \sigma_0) t}$ to obtain 
\begin{eqnarray}
   i \hbar \, \partial_t w =
  - (\hbar^2/2m) \partial_z^2 w +
 \hbar \omega_{\perp} 2 a (|w|^2 - \sigma_0) w.
\label{gp42w}
\end{eqnarray}
Equation (\ref{gp42w}) has the standard (quadratic) nonlinear term and leads 
to a speed of sound, $c_w$, which satisfies the equation $m c_w^2 = 2 \hbar 
\omega_{\perp} \sigma_0 a$ \cite{JKP} [see also Eq.\,(\ref{bog}).] Since 
$\sigma_0 = n_0 \pi a_{\perp}^2$, we see that $m c_w^2 = n_0 U_0/2$.

Writing $w=\sqrt \sigma e^{i \phi}$ and separating the real and imaginary 
parts of Eq.\,(\ref{gp42w}), we obtain the two hydrodynamic equations 
\begin{eqnarray}
     \frac {\hbar^2} {2m}
   \left( \frac {\partial \sqrt \sigma} {\partial z} \right)^2 =
  \left( 2 \hbar \omega_{\perp} \sigma a - m u^2 \right)
 \frac {(\sigma - \sigma_0)^2} {2\sigma},
\label{Eul}
\end{eqnarray}
\begin{eqnarray}
v = u (1 - \sigma_0 / \sigma).
\label{vel}
\end{eqnarray}
Here, we have imposed the boundary condition $v \to 0$ for $\sigma \to 
\sigma_0$. The solution of Eq.\,(\ref{Eul}) is
\begin{eqnarray}
   \sigma(z)/\sigma_0 - 1 = - \frac {\cos^2 \theta}
{\cosh^2(z \cos \theta /\zeta)},
\label{solEull}
\end{eqnarray}
where $\theta = {\rm arcsin}(u/c_w)$ and $\zeta=2\xi(n_0)$ with $\xi(n_0)$ 
equal to the coherence length for $n_0 = \sigma_0/ (\pi a_{\perp}^2)$ (i.e., 
$\zeta = a_{\perp}/ (2 \sigma_0 a)^{1/2}$).  The wavefunction $w$ can also 
be written as 
\begin{eqnarray}
w = \sqrt{\sigma_0} \, [ i \sin \theta + \cos \theta
 \tanh ( z \cos \theta / \zeta) ].
\label{newfwf}
\end{eqnarray}

{\it Energy and momentum of the solitary wave.}
In the limit of weak interactions, Eq.\,(\ref{gp42w}) implies that
\begin{eqnarray}
   {\cal E} = \int \left( \frac {\hbar^2} {2 m}
  \frac {\partial w^*} {\partial z} \frac {\partial w}
 {\partial z} + \hbar \omega_{\perp} a
(w w^*)^2 - 2 \hbar \omega_{\perp} a \sigma_0 w w^*
\right. \nonumber \\ \left.
+ \hbar \omega_{\perp} a \sigma_0^2 \right) dz,
\label{ew1}
\end{eqnarray}
where the final term, which represents the energy of the background density 
of atoms, ensures convergence of the integral.  Equation (\ref{ew1}) can be 
written as
\begin{eqnarray}
        {\cal E} = \int \left[ \frac {\hbar^2} {2 m}
      \left( \frac {\partial \sqrt \sigma} {\partial z} \right)^2
     + \frac {\hbar^2 \sigma} {2 m}
    \left( \frac {\partial \phi} {\partial z} \right)^2
   \right. \nonumber \\ \left.
  +\hbar \omega_{\perp} a (\sigma - \sigma_0)^2
 \right] dz.
\label{ew2}
\end{eqnarray}
Since $v=\hbar \partial_z \phi/m$, Eqs.\,(\ref{Eul}) and (\ref{vel})
allow us to reduce Eq.\,(\ref{ew2}) to the simpler form
\begin{eqnarray}
  {\cal E} = 2 \hbar \omega_{\perp} a \int (\sigma - \sigma_0)^2 \, dz.
\label{ew2f}
\end{eqnarray}
Given the solitary wave profile of Eq.\,(\ref{solEull}), Eq.\,(\ref{ew2f}) 
yields
\begin{eqnarray}
  {\cal E} = 
   (4 \sqrt 2/3) {\cal E}_0
  \cos^3 \theta,
\label{ew2fs}
\end{eqnarray}
where ${\cal E}_0=\hbar \omega_{\perp} (\sigma_0 a)^{1/2} \sigma_0 a_{\perp}$. 

Calculation of the momentum $\cal P$,
\begin{eqnarray}
  {\cal P} = - i \hbar \int w^* \frac {\partial w} {\partial z} \, dz
 = m \int \sigma(z) v(z) \, dz,
\label{pst1}
\end{eqnarray}
requires a more careful description of the solitary-wave profile for 
large $|z|$.  This is most easily accomplished by requiring that the 
solitonic wavefunction is periodic on a large interval $[-L/2 , +L/2]$.  
The solution to this problem for all $L$ can be expressed analytically in 
terms of Jacobi elliptic 
functions \cite{Tsuzuki}.  For our purposes, it is sufficient to note 
the behaviour of $\sigma$ and $\phi$ in the limit of large $L$.  
For $z \sim {\cal O}(L^0)$, $\sigma$ is still given by Eqs.\,(\ref{solEull}) 
and (\ref{newfwf}).  For positive $u$, Eq.\,(\ref{vel}) indicates that
$\partial_z \phi$ is negative and that a net phase will accumulate as we move 
from the center of the solitary wave at $z=0$ towards $z = +L/2$.  For $z \sim 
{\cal O}(L)$, however, the periodic $\sigma$ approaches a constant value 
larger than $\sigma_0$ by an amount proportional to $1/L$.  In this region, 
$\partial_z \phi$ is positive, the total phase accumulated between $z=0$ 
and $+L/2$ is precisely $0$, and the periodic boundary conditions are 
fulfilled.  Evidently, this large-$z$ behaviour makes a finite contribution 
to the momentum in the $L \to \infty$ limit which is readily evaluated from 
Eq.\,(\ref{pst1}).  To leading order, the 
momentum can be evaluated using Eqs.\,(\ref{solEull}) and (\ref{newfwf}) to 
obtain
\begin{eqnarray}
 {\cal P} = m \int_{-\infty}^{\infty} (\sigma - \sigma_0) v(z) \, dz
= m u \int_{-\infty}^{\infty} \frac {(\sigma-\sigma_0)^2} {\sigma} \, dz.
\label{pst2a}
\end{eqnarray}
Using the order parameter of Eq.\,(\ref{newfwf}), we find that
\begin{eqnarray}
  {\cal P} =
   {\cal P}_0 ( \pi u / |u|
  - 2 \theta - \sin 2 \theta),
\label{pst133344}
\end{eqnarray}
where ${\cal P}_0 = \sigma_0 \hbar$. The momentum ${\cal P}$ was also
determined in Ref.\cite{J} using the macroscopic relation ${\cal P}=\int 
u^{-1} (\partial {\cal E}/\partial u) \, du$. This yields a result identical 
to Eq.\,(\ref{pst133344}). Equation (\ref{pst133344}) implies that a maximum 
momentum of ${\cal P}_{\rm max}=\pi\sigma_0\hbar$ is obtained for $u=0$. Note 
that, to leading order in $1/L$, the energy of Eq.\,(\ref{ew2fs}) is 
unaltered by the imposition of periodic boundary conditions.  

Combining Eqs.\,(\ref{ew2fs}) and (\ref{pst133344}), we arrive at the 
dispersion relation ${\cal E}({\cal P})$ for the Lieb mode in the limit of 
weak interactions.  The solid line of Fig.\,1(a) shows this result.  
For ${\cal P} \to 0$, 
${\cal E} = c_w {\cal P}$, in agreement with the usual Bogoliubov dispersion 
relation discussed below.  The Lieb mode terminates at ${\cal P}= 
{\cal P}_{\rm max}$ where it has an energy of ${\cal E}/{\cal E}_0
=4 \sqrt{2}/3$. 

{\it Bogoliubov mode in the weak interaction limit}.
Equation (\ref{gp42w}) implies that the Bogoliubov mode obeys the dispersion 
relation
\begin{eqnarray}
    \frac {\cal E} {{\cal E}_0}
  = \sqrt 2 \frac {|{\cal P}|} {{\cal P}_0}
 \sqrt{ 1 + \left( \frac {\cal P} {{\cal P}_0} \right)^2
\frac {\sigma_0 a} 8 \left( \frac {a_{\perp}} {a} \right)^2 }.
\label{bog}
\end{eqnarray}
Choosing $\sigma_0 a = 0.1$, $a_{\perp} = 1$ $\mu$m, 
and $a = 100$ \AA, the coefficient inside the square root is 125.
This number is relatively large because the characteristic 
wavevector corresponding to ${\cal P}_0$, i.e., $\sigma_0$, is much larger 
than the inverse coherence length, $\xi^{-1}(n_0)$.  
Specifically, $\xi^{-1}(n_0) \sim (8 
\sigma_0 a)^{1/2} / a_{\perp}$.  Thus, $\sigma_0 \xi(n_0) \sim a_{\perp} 
(\sigma_0/a)^{1/2} \sim 10$.  The dotted line in Fig.\,1(a) shows the 
dispersion relation of Eq.\,(\ref{bog}) for the above numbers.

{\it The strong coupling limit}.
In the limit of strong interactions, $n_0 U_0 \gg \hbar \omega_{\perp}$,
we use the Thomas-Fermi approximation for the transverse profile,
$|g|^2 = 2 ( \pi R_{\perp}^2 )^{-1} ( 1 - \rho^2 / R_{\perp}^2 )$,
with $R_{\perp}/ a_{\perp} = 2 (\sigma a)^{1/4}$ \cite{KP}.  From 
Ref.\cite{JKP} we again see that $w=f e^{-2i \omega_{\perp} 
(\sigma_0 a) t}$ satisfies the equation
\begin{eqnarray}
  i \hbar \, \partial_t w = - (\hbar^2/2m)
 \partial_z^2 w + 2 \hbar \omega_{\perp} a^{1/2}
(|w| - |w_0|) w.
\label{gp41rev}
\end{eqnarray}
The effective equation obeyed by $w$ now involves a modified non-linear 
term proportional to $|w|$.  Equation (\ref{gp41rev}) implies a sound speed 
of $m c_s^2 =\hbar \omega_{\perp} (\sigma_0 a)^{1/2}$ (see also 
Eq.\,(\ref{last}).]  Since $n_0 U_0 = 2 \hbar \omega_{\perp} 
(\sigma_0 a)^{1/2}$, this becomes $m c_s^2 = n_0 U_0/2$ \cite{JKP,KP}.

Using Eq.\,(\ref{gp41rev}), we again write $w=\sqrt \sigma e^{i \phi}$
and separate real and imaginary parts.  The velocity is still
given by Eq.\,(\ref{vel}). In addition, Eq.\,(\ref{gp41rev}) 
implies that
\begin{eqnarray}
     \frac {\hbar^2} {2m}
   \left( \frac {\partial \sqrt \sigma} {\partial z} \right)^2 =
   \frac 2 3 \hbar \omega_{\perp} a^{1/2}
  (2 \sigma^{3/2} &-& 3 \sigma_0^{1/2} \sigma + \sigma_0^{3/2})
\nonumber \\
 &-& m u^2 \frac {(\sigma - \sigma_0)^2} {2\sigma}.
\label{Eul1}
\end{eqnarray}
This equation provides a relation between $u$ and the
minimum value of $\sigma$, $\sigma_{\rm min}$.  For a given
$u$, $\sigma_{\rm min}$ is given by the non-trivial root ($\sigma \neq
\sigma_0$) of the right side of Eq.\,(\ref{Eul1}).

{\it Energy and momentum in the strong interaction limit.}
Using Eq.\,(\ref{pst2a}) we find that the momentum is given by 
\begin{eqnarray}
   {\cal P}/{\cal P}_0 =   u / {c_s}
  \int  (y-1)^2/y \, \, dx,
\label{feq1}
\end{eqnarray}
where $x=z (\sigma_0 a)^{1/4}/a_{\perp}$ and $y = \sigma/\sigma_0$.  Here, 
$\sigma$ is the solution of Eq.\,(\ref{Eul1}).  In addition, 
Eq.\,(\ref{gp41rev}) gives an energy
\begin{eqnarray}
  {\cal E} = \int \left[ \frac {\hbar^2} {2 m}
 \left( \frac {\partial \sqrt \sigma} {\partial z} \right)^2
  + \frac {\sigma \hbar^2} {2 m}
  \left( \frac {\partial \phi} {\partial z} \right)^2 + \phantom{XXX}
\right. \nonumber \\ \left.
 +  \frac 2 3 \hbar \omega_{\perp} a^{1/2}
\left(2 \sigma^{3/2} - 3 \sigma \sigma_0^{1/2}
+ \sigma_0^{3/2} \right)
\right] dz.
\label{ew22}
\end{eqnarray}
The final term in the integral again guarantees its convergence and 
corresponds to the energy of the background density.

Since $v=\hbar \partial_z \phi/m$, Eq.\,(\ref{Eul1}) and the formula 
$v = u (1 - \sigma_0 / \sigma)$ allow us to write Eq.\,(\ref{ew22}) as
\begin{eqnarray}
  {\cal E} = (4/3) \hbar \omega_{\perp} a^{1/2} \int
(2 \sigma^{3/2} - 3 \sigma \sigma_0^{1/2} + \sigma_0^{3/2}) \, dz.
\label{ennn1}
\end{eqnarray}
Introducing the unit of energy ${\cal E}_0^{'} = \hbar \omega_{\perp}
(\sigma_0 a)^{1/4} \sigma_0 a_{\perp}$, 
\begin{eqnarray}
   {\cal E} / {\cal E}_0^{'} = \frac 4 3 \int
(2 y^{3/2} - 3 y + 1) \, dx.
\label{feq10}
\end{eqnarray}

We have solved Eq.\,(\ref{Eul1}) numerically to obtain $\sigma(z)$ for 
various values of $u$.  This numerical solution was used in 
Eqs.\,(\ref{feq1}) and (\ref{feq10}) to determine ${\cal P}(u)$ and 
${\cal E}(u)$, respectively.  The solid line in Fig.\,1(b) shows the resulting 
dispersion relation for the Lieb mode in the limit of strong interactions.  
The $x$ axis is measured in units of ${\cal P}_0$, and the $y$ axis is 
measured in units of ${\cal E}_0^{'}$. As in the case of weak interactions,
the slope of the curve for small values of ${\cal P}$ is equal to $c_s$,
and also the curve terminates at $\pi {\cal P}_0$ with ${\cal E} \approx 1.5 
\, {\cal E}_0^{'}$.

{\it Bogoliubov mode in the limit of strong interactions}.
Equation (\ref{gp41rev}) implies that the dispersion relation associated
with the Bogoliubov mode is 
\begin{eqnarray}
    \frac {\cal E} {{\cal E}_0^{'}}
= \frac {|{\cal P}|} {{\cal P}_0}
 \sqrt{ 1 + \left( \frac {\cal P} {{\cal P}_0} \right)^2
\frac 1 4 (\sigma_0 a)^{3/2} \left( \frac {a_{\perp}} a \right)^2  }.
\label{last}
\end{eqnarray}

For $\sigma_0 a = 1$, $a_{\perp} = 1$ $\mu m$,
and $a = 100$ \AA, the coefficient inside the square root 
is $\approx 2500$, which is $\gg 1$, since the characteristic wavevector 
corresponding to ${\cal P}_0$, i.e., $\sigma_0$, is $\gg$ than
$\xi^{-1}(n_0)$: $\sigma_0 \xi(n_0) 
\sim \sigma_0 a_{\perp}/(\sigma_0 a)^{1/4}$. The dotted line in Fig.\,1(b)
shows the dispersion relation of Eq.\,(\ref{last}) for this choice of 
parameters. Once
again, the Lieb and the Bogoliubov modes have identical dispersion relations 
in the limit of long wavelength excitations, as expected. 

The methods adopted here are expected to be reliable for weak interactions 
\cite{JKP} for which the width of the solitary wave is much larger than the 
transverse width of the cloud.  In this regime the transverse degrees of 
freedom are frozen out, and the behavior of the system is essentially 
one-dimensional.  As the strength of the interaction is increased, 
it has been demonstrated in Ref.\,\cite{MHS} that the dark solitary waves 
become unstable.  (See also Ref.\,\cite{Feder}.)  As shown in this reference, 
dark solitary waves in a cylindrical trap become unstable for $n_0 U_0/\hbar 
\omega_{\perp} \ge 2.4$.  Assuming that the system is in the Thomas-Fermi
regime, $\sigma_0 / (\pi R_{\perp}^2) = n_0/2$, with $R_{\perp}^2 = 4
a_{\perp}^2 (\sigma_0 a)^{1/2}$ \cite{KP}, and therefore the corresponding 
critical value of $\sigma_0 a$ is $\approx 1.2^2 = 1.44$.  Thus, our 
approach should also provide a reasonable variational description of 
the dispersion relation in the strong interaction regime provided only 
that dark solitary waves are stable.  In addition, we note that the 
Bogoliubov dispersion relation obtained here is in analytic agreement with 
numerical solutions \cite{SF} in the large and small momentum limits for 
both weak and strong interactions.  Differences of only a few percent for 
intermediate momenta are found for $\sigma_0 a$ as large as 1. 

In summary, we have calculated the dispersion relation ${\cal E}=
{\cal E}({\cal P})$ of a sound wave and of a solitary wave in a 
quasi one-dimensional Bose-Einstein condensate of atoms confined in 
a harmonic, cylindrical trap in the limits of weak and strong 
interactions.  For solitary waves, in both limits the spectrum has the 
same qualitative behaviour: it is linear for long wavelength excitations and
coincides with the Bogoliubov mode. For shorter wavelength excitations, 
it lies below the Bogoliubov mode, and it terminates at a maximum momentum for
which the group velocity vanishes.
\vskip0.1pc 
We would like to thank N. Papanicolaou for pointing out the existence
of the Lieb mode to us and for useful discussions. We also acknowledge
profitable discussions with C. J. Pethick. G.M.K. was supported  
by the Swedish Research Council (VR), and by the Swedish Foundation for
Strategic Research (SSF). 
G.M.K. would like to thank the Physics Dept.
of the Univ. of Crete for its hospitality.

\noindent
\begin{figure}
\begin{center}
\epsfig{file=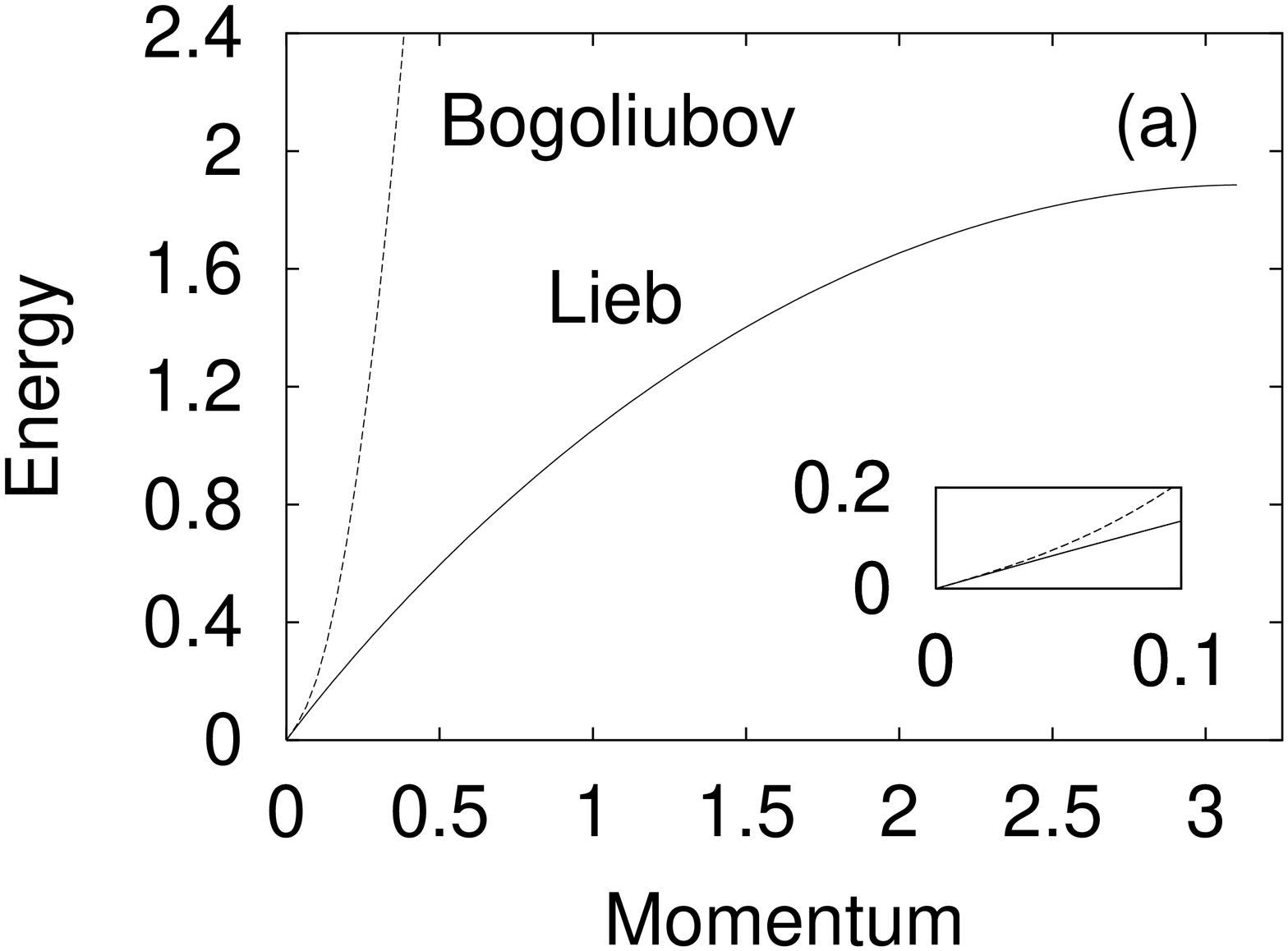,width=4.2cm,height=6.0cm,angle=-90}
\\
\epsfig{file=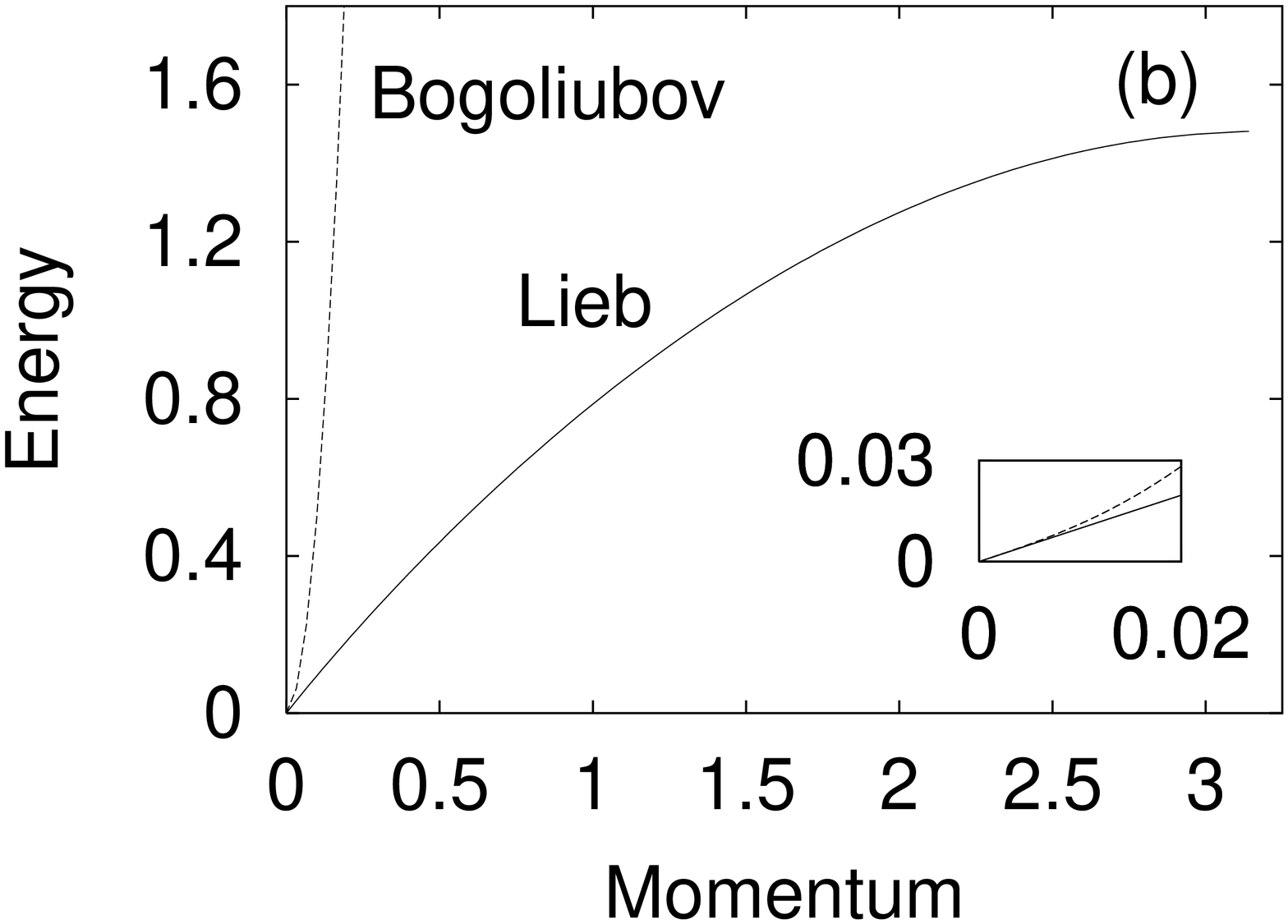,width=4.2cm,height=6.0cm,angle=-90}
\vskip0.5pc
\begin{caption}
{(a) Dispersion relation ${\cal E} = {\cal E} ({\cal P})$
in the limit of weak interactions for a solitary wave (solid curve) 
and for the Bogoliubov mode (dotted line.) The energy is measured in units 
of ${\cal E}_0$ and the momentum in units of ${\cal P}_0$. (b) Same as (a)
in the limit of strong interactions, with the energy measured in 
units of ${\cal E}_0^{'}$. The inset shows the same graph on a smaller scale,
for long-wavelength excitations.}
\end{caption}
\end{center}
\label{FIG1}
\end{figure}
\noindent

\end{document}